# IMPACT OF CHANGING THE WET DEPOSITION SCHEMES IN LDX ON 137-CS ATMOSPERIC DEPOSITS AFTER THE FUKUSHIMA ACCIDENT

*Arnaud Quérel[1], Denis Quélo[1], Yelva Roustan[2], Anne Mathieu[1], Mizuo Kajino[3], Thomas Sekiyama[3], Kouji Adachi[3], Damien Didier[1], Yasuhito Igarashi[3], Takashi Maki[3]*

[1] IRSN, PRP-CRI, SESUC, BMCA, Fontenay-aux-Roses, France
[2] CEREA, ENPC/EDF R&D, Champs-sur-Marne, France
[3] MRI, Japan Meteorological Agency, Tsukuba, Japan

**Abstract**: The Fukushima-Daiichi release of radioactivity is a relevant event to study the atmospheric dispersion modelling of radionuclides. Actually, the atmospheric deposition onto the ground may be studied through the map of measured Cs-137 established consecutively to the accident. The limits of detection were low enough to make the measurements possible as far as 250km from the nuclear power plant. This large scale deposition has been modelled with the Eulerian model ldX. However, several weeks of emissions in multiple weather conditions make it a real challenge. Besides, these measurements are accumulated deposition of Cs-137 over the whole period and do not inform of deposition mechanisms involved: in-cloud, below-cloud, dry deposition.
In a previous study (Quérel et al., 2016), a comprehensive sensitivity analysis was performed in order to understand wet deposition mechanisms. It has been shown that the choice of the wet deposition scheme has a strong impact on assessment of deposition patterns. Nevertheless, a "best" scheme could not be highlighted as it depends on the selected criteria: the ranking differs according to the statistical indicators considered (correlation, figure of merit in space and factor 2). A possibility to explain the difficulty to discriminate between several schemes was the uncertainties in the modelling, resulting from the meteorological data for instance. Since the move of the plume is not properly modelled, the deposition processes are applied with an inaccurate activity concentration in the air. In the framework of the SAKURA project, an MRI-IRSN collaboration, new meteorological fields at higher resolution (Sekiyama et al., 2013) were provided and allow to reconsider the previous study.
An update including these new meteorology data is presented. In addition, the focus is put on the deposition schemes commonly used in nuclear emergency context.

***Key words:*** *Fukushima; sensitivity analysis; wet deposition, atmospheric dispersion model*

## INTRODUCTION
Wet deposition processes are a crucial component of radionuclide atmospheric transport and soil contamination models. Within atmospheric dispersion model these processes are represented by coefficients that quantify the proportion of scavenged radionuclides per unit of time. Radionuclides are captured by hydrometeors (droplets, snowflakes, etc...) and brought to the Earth's surface by precipitations (rain, snow, etc...). If the capture occurs when the condensed water is aloft in the atmosphere, the phenomenon is called rainout (or in-cloud scavenging). If it occurs during the precipitations, it is called washout (or below-cloud scavenging). These processes are physically separated and can be distinctly represented with their own scavenging coefficients. A wide range of parametrizations are proposed in the scientific literature to determine these coefficients (Duhanyan and Roustan, 2011), and currently no scientific consensus allows to discriminate between them.
Furthermore, the Fukushima case shows significant episodes of wet deposition, making it particularly well suited to study the wet deposition modelling. The release of long-lived species 137-Cs, bound to atmospheric particles, is simulated with the Eulerian long-range dispersion model from the IRSN, ldX (Groëll et al., 2014) .

In a previous step, a sensitivity analysis, relying on an ensemble-type approach, was performed in an attempt to discriminate between the various wet deposition schemes proposed in the literature (Quérel et al., 2016). The results of this analysis confirm the strong sensitivity of the simulated deposits to the choice of deposition schemes, both in terms of magnitude and spatial patterns and similar results were shown in Leadbetter et al. (2015). However it was not possible to identify a particular scheme which lead to an overall improvement in modelling performance and call for caution regarding the results obtained. This could be partly attributed to the uncertainties remaining in the meteorological fields and in the source terms used to drive the atmospheric dispersion model. To go further, finer meteorological data and more up-to-date source terms are now considered. Moreover, since ldX aims at nuclear emergency response, it is suitable to have at our disposal in our modelling framework other parameterizations for wet deposition, like the ones used in similar operational atmospheric dispersion models.

**WET DEPOSITION SCHEMES**
In addition to the IRSN ones, other wet deposition schemes are considered: those documented in the WMO Task Team report investigating the impact of meteorology on the dispersion of radioactive material from Fukushima-Daiichi Nuclear Power Plant (Draxler et al., 2012). They are issued from the following atmospheric dispersion models: **CMC-MLDP0** (D'Amours et al., 2010), **HYSPLIT** (Draxler and Hess, 1997), **NAME** (Jones et al., 2007), **RATM** (Shimbori et al., 2010) and **FLEXPART** (Stohl et al., 2010). The description of deposition schemes are reported in **Table 1**.

**Table 1:** Deposition schemes used by models involved in the WMO Task Team.

| Atmospheric transport modelling | Below-cloud scheme | In-cloud scheme |
|---|---|---|
| CMC-MLDP0 | $\Lambda = 0$ | $\Lambda = 3 \times 10^{-5}$ |
| FLEXPART | $\Lambda = 10^{-5} I^{0.8}$ | Hertel et al. (1995) |
| HYSPLIT | $\Lambda = 10^{-6}$ | Hertel et al. (1995) with $S = 4 \times 10^4$ |
| IRSN | $\Lambda = 5 \times 10^{-5} I$ | $\Lambda = 5 \times 10^{-5} I$ |
| NAME | $\Lambda = 8.4 \times 10^{-5} I^{0.79}$ | $\Lambda = 3.36 \times 10^{-4} I^{0.79}$ |
| RATM | $\Lambda = 2.78 \times 10^{-5} I^{0.75}$ | Hertel et al. (1995) with LWC model |

**METHODOLOGY AND INPUT DATA**
In this study, washout and rainout parameterizations are not considered independently in our ensemble-type approach. Then, model configurations look through a pair of deposition schemes (listed in table1), in addition to meteorological data, precipitation field and source terms. Input data considered are listed below:
- Meteorological data: number 1 and number 8 of Sekiyama's ensemble (Sekiyama et al., 2013).
- Precipitations: issued from the meteorological simulation or radar rainfall data corrected by rain-gauges observations (Saito et al., 2015).
- Source term: Katata (2015), Saunier (2013) and Terada (2012).

All simulations are performed using the long-range transport model ldX with common settings except for the wet scavenging schemes. Clouds vary between fixed altitudes of 800 m and 5000 m. Dry deposition is modelled through a constant deposition velocity of $2 \times 10^{-3}$ m.s$^{-1}$. Horizontal grid comes from the meteorological data whose resolution is 0.03° and vertical resolution follows a non-linear scale ranging from 0 to 5546 m (0, 40, 85, 141,… 5546 m).

Simulations are compared to a data set of 137-Cs deposit observations which comprises airborne measurements made by the US Department of Energy as well as ground measurements collected by the Ministry of Education, Culture, Sports, Science and Technology of Japan. This data are averaged onto the Model grid (see Figure 1, a).

**DATA EXPLOITATION**
Three statistical indicators are used to evaluate the performance of simulations to reproduce the deposits:
- Pearson correlation, used to quantify the linear relationship between observations and simulations.
- Factor 2, used to obtain an overall measure of the magnitude of discrepancies between observations and simulations, with equal weighting of low and high values.
- Figure of Merit in Space (FMS), used to compare simulated and observed deposition patterns.

A minimum value of $10 kBq.m^{-2}$ is applied for observed deposits.

Influence of a specific scheme rather than others is investigated within this ensemble type of simulations. Issues considered are: How many simulations are improved by this given choice? Is it a significant improvement? In order to estimate this impact, each simulation performed with the chosen deposition schemes is compared to simulations sharing the same configuration. This comparison is made for each statistical indicator is a distribution of discrepancies for each choice of deposition schemes.

These obtained distributions are then drawn with the help of whisker box.

**RESULTS**
Figure 1 shows maps of deposits considering different wet deposition schemes. In this example, NAME schemes lead to a greater deposit than IRSN's one.

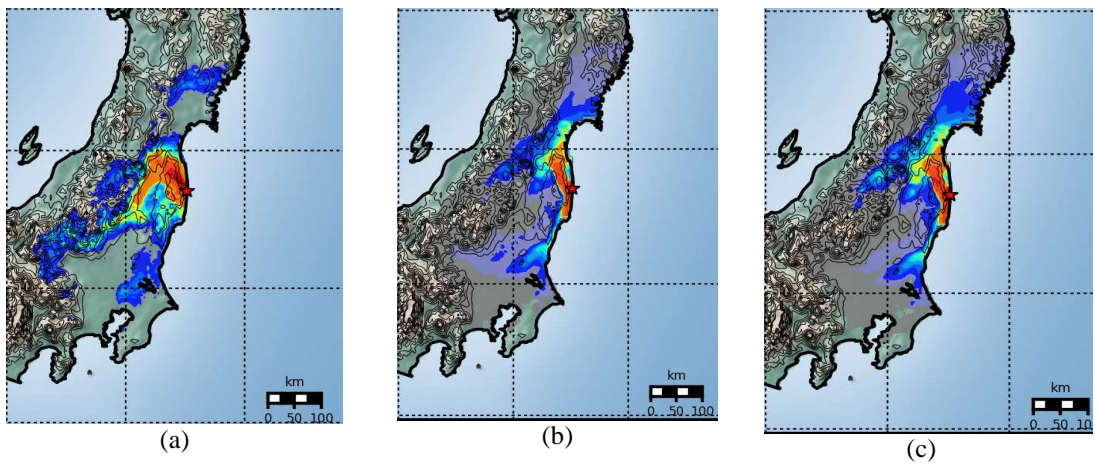

(a)    (b)    (c)

**Figure 1**: Deposit (a) Observed; (b) simulated using IRSN deposition schemes; (c) simulated using NAME deposition schemes. Both simulations are performed with Katata source term, Sekiyama n°8 meteorological data and radar rainfall data.

Figure 2 shows an example of the impact on statistical indicators using the possible options for wet deposition. A sensitivity of a given choice is observed on all the indicators. Some wet deposition schemes appear to give better results than others and one can wonder if it is specific to this case study or to the input data (meteorological fields and source terms) included in this ensemble-type approach. These preliminary results need to be consolidated.

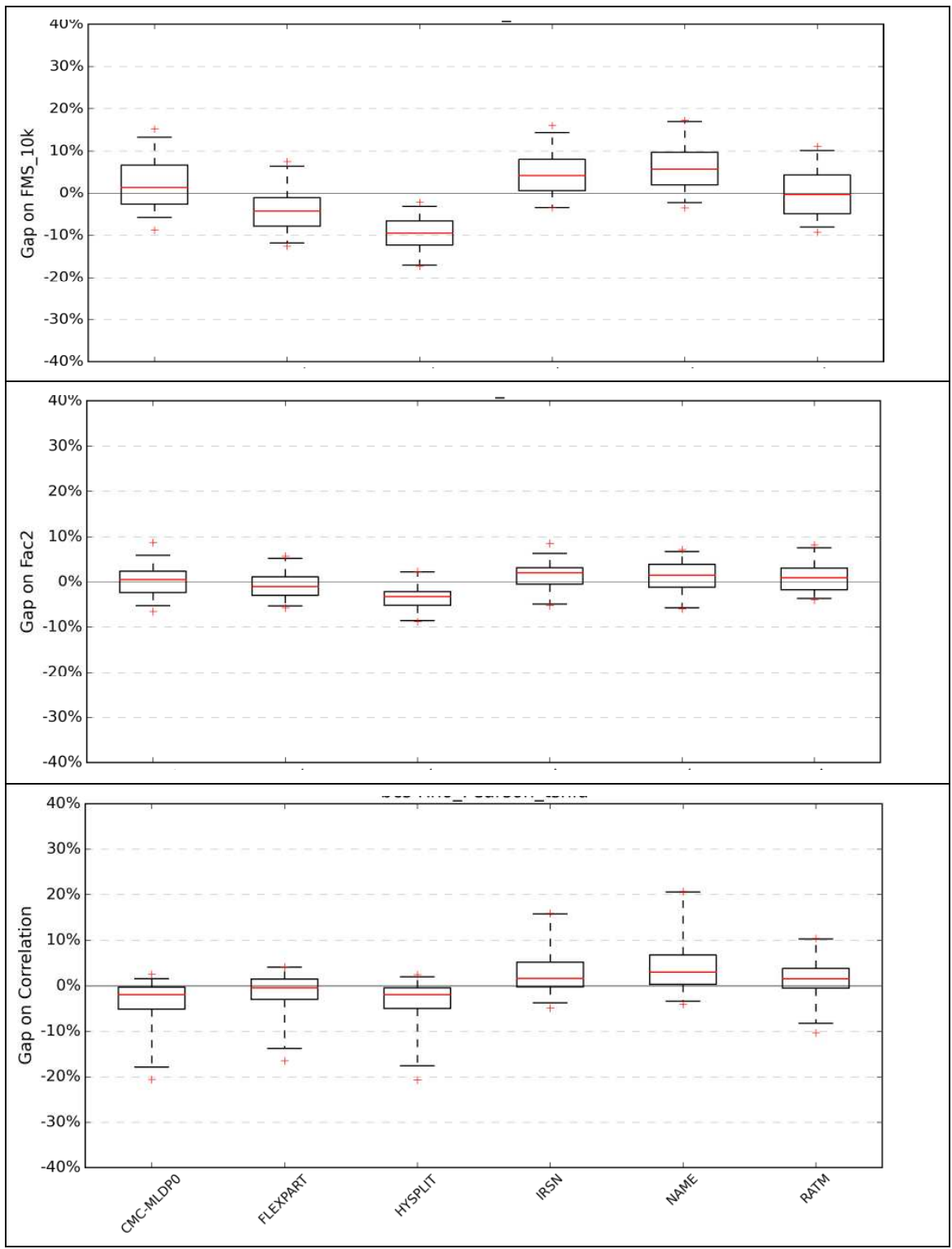

**Figure 2**: Impact of wet deposition schemes on FMS, factor 2 and correlation.